# Coherent control of acoustic phonons in a silica fiber using a multi-GHz optical frequency comb

Mamoru Endo[1], Shota Kimura[1], Shuntaro Tani[1] and Yohei Kobayashi[1,*]

[1]*The Institute for Solid State Physics, The University of Tokyo, 5-1-5 Kashiwanoha, Kashiwa, Chiba 277-8581, Japan*

**Multi-gigahertz mechanical vibrations stemming from interactions between light fields and matter—also known as acoustic waves or phonons—have long been a subject of study. In recent years, for the purpose of the next generation of high-speed telecommunication applications, specially designed functional devices have been developed to enhance the light-matter interaction strength, since the excitation of acoustic phonons by a continuous wave (cw) laser alone is insufficient. However, with such structure-dependent enhancements, the strength of the interaction cannot be aptly and instantly controlled. We propose a new technique to control the effective interaction strength, which is not via the material structure in the spatial domain, as with the above-mentioned specially designed functional devices, but through the structure of light in the time domain.**

**Here we show the effective excitation of acoustic phonons in a single mode fiber using our newly developed optical frequency comb (OFC). In addition to controlling the time structure of the OFC via its repetition frequency, we demonstrate coherent control over acoustic phonons by exciting, enhancing, and suppressing them. We believe this work to represent an important step towards "comb-matter interactions", that it will be useful in topical areas of research such as the investigation of optically-accessible magnons and**

**skyrmions, and that it will lead to increasingly flexible tools in combination with the aforementioned functional devices.**

Research on the interaction between light fields and acoustic phonons in lattice media has greatly advanced over the past decade. It holds enormous potential in scientific studies and practical areas, where possible applications include the generation of ultra-low noise lasers[1], optical frequency combs (OFCs)[2,3], coherent control[4,5], optical storage[6–8], slow and fast light generation[9], optical switching[10], and Brillouin cooling[11], to name of few. This light-matter interaction is caused by a permittivity change of a medium by an incident electric field. Due to the weak nature of the interaction, specially-designed photonic crystal fibers[5,12–14], suspended waveguides[15,16] or microresonators[1,6], are devised to constrain both the optical fields and the acoustic modes, to improve the interaction strength. However, with such structure-dependent interaction enhancement, there is a lack of ability to control the interaction strength dynamically or rapidly. In optical fibers, Brillouin scattering is the main process involved in the generation of acoustic phonons and it requires long interaction lengths, like with cw or long pulsed (at least few nanoseconds)[17] lasers. Long pulses, however, are unsuitable for telecommunication applications that require fast and efficient control of the light-matter interaction.

We demonstrate here that if a second pulse is launched after a first one well within the lifetime of a given acoustic phonon, this acoustic phonon can be coherently enhanced or damped. To achieve this, we used the picosecond pulses from our own developed multi-GHz OFC which has its repetition rate set to the phonon resonance frequency (up to 16 GHz)[18]. Enhancing and damping can be controlled by the pulse repetition rate or the time delay between the pulses, which make multi-GHz OFCs suitable for the effective and coherent excitation of acoustic phonons.

Brillouin scattering results from photon-phonon interactions through the electrostriction effect[19]. When an acoustic phonon satisfies the phase-matching condition, it scatters light in the counter-propagating backward direction (BWD). This is shown in Fig. 1 (a), as a cw (or long pulsed) laser beam propagates in the forward direction (FWD) with an angular frequency of $\omega_{FWD}$ (FWD optical, red line) and is coupled into an optical fiber, the beam is scattered by the thermally excited acoustic phonons in the fiber and the back-scattered component at the down-shifted Stokes frequency, $\omega_{BWD}$ (BWD optical, purple line), is created. The density fluctuations caused by the two counter-propagating optical fields create a traveling acoustic wave with an angular frequency $\Delta\omega = \omega_{FWD} - \omega_{BWD}$ and a speed of $\frac{\Delta\omega}{n_{FUT}\omega_{FWD}}$, where $n_{FUT}$ represents the refractive index of the fiber-under-test (FUT). This acoustic wave (FWD acoustic, sky blue line) scatters the FWD optical beam to create the BWD scattered light again and thus the cycle continues. When the speed of the FWD acoustic wave matches the speed of sound in the fiber ($v_A$), stimulated scattering occurs. The phase-matching condition is expressed as

$$\omega_{AP} = \Delta\omega = 2\pi \times \frac{2n_{FUT}v_A}{\lambda_{FWD}}, \tag{1}$$

where $\lambda_{FWD}$ represents the wavelength of the FWD optical wave and the concept is described in the dispersion diagram of Fig. 1 (b). In other words, the acoustic wave leads to a Brillouin gain and loss at the center frequencies of $\omega_{FWD} - \omega_{AP}$ and $\omega_{FWD} + \omega_{AP}$, respectively, as shown in the gain profile in Fig. 1(c). The BWD light has a broad spectrum due to Doppler frequency shifts caused by the thermally excited acoustic phonons. The main component at $\omega_{FWD}$ is referred to as Rayleigh scattering and the components surrounding this main central peak are called Rayleigh wings. The Brillouin gain amplifies the minuscule Stokes component in the Rayleigh wing at $\omega_{FWD} - \omega_{AP}$. Under these conditions, the photon-phonon interaction strength is weak and long interaction lengths or specially designed structures are needed to enhance it. The typical frequency of acoustic phonons in a silica fiber is approximately 16 GHz for a pump light at a wavelength of 1 μm.

Figures 1 (d-f) are equivalent to 1 (a-c), respectively, only they relate to the case where the pump beam is a train of pulses, like that of an OFC with a repetition frequency of $\omega_{rep}/(2\pi)$. The number of the longitudinal modes is $N$ and the optical angular frequency of the $n$-th mode is $\omega_{FWD}^{(n)} = \omega_{FWD}^{(1)} + (n-1)\omega_{rep}$. More specifically, Fig. 1 (d-f) depict the instance when the repetition frequency is equal to the resonance frequency. This leads to the backscattered component at the Stokes angular frequency of the first longitudinal mode, $\omega_{FWD}^{(1)} - \omega_{AP}$, to have a gain spectrum similar to the one depicted in Fig. 1(c). However, as shown in Fig. 1(f), the potential of this technique comes in the gain spectra of the other longitudinal modes. The Brillouin gain of the next longitudinal mode at $\omega_{FWD}^{(1)}$ coherently adds to the center of the Rayleigh scattering of the next longitudinal mode at $\omega_{FWD}^{(2)}$, where the amplitude is much higher than at the Rayleigh wings, and a markedly enhanced BWD component is produced. At the Stokes frequency of all the other longitudinal modes, the Brillouin loss and gain cancel each other out and this through to the anti-Stokes frequencies of the two highest longitudinal modes of the OFC, where for the first of the two, the Brillouin loss attenuates the center of the Rayleigh scattering of the last longitudinal mode at $\omega_{FWD}^{(N)}$, and in the case of the second one of the two, the scenario is that depicted at the anti-Stokes frequency in Fig. 1(c).

It should be noted that the phonon frequency depends on the optical frequency of the pump light, as shown with equation (1), and the center of the backscattered Rayleigh, Stokes and anti-Stokes gains are not perfectly matched over all the longitudinal modes of the OFC. The profile of the Brillouin gain in conventional single mode fibers at room temperature has a spectral width of several tens of MHz and the phase matching condition is more relaxed for pump light with spectral widths of sub-hundred GHz (corresponding to a pulse duration of several tens of ps).

Figure 2 (a) illustrates the experimental apparatus and the details can be found in the Method. To observe a dependence between the repetition rate of the OFC and the acoustic phonons, we used an OFC with a repetition rate of 15.6 GHz which produced a 30-ps pulse train with a maximum average power of 100-mW. The optical pulses were coupled into a 10-km FUT with a core diameter of 11 μm (FutureGuide SR-15E, Fujikura) after an optical circulator (CIR). The input and BWD scattered lights were measured with an optical spectrum analyzer (OSA) and a power meter (PM) through a 99:1 fiber coupler and an optical switch (OS). The input is used as the reference (ref.), and the BWD light is used to measure the phonon amplitude (meas.). Figure 2 (b) represents the phonon amplitude estimated from the power of the BWD light as a function of the OFC's repetition frequency for an input power of 20 mW. The black dots and the error bars are the measured values, and the red solid line is the Gaussian fit. The BWD power was strongly enhanced when the repetition rate matched the phonon resonance (15.57 GHz) within the phonon linewidth (approximately 40 MHz in our case). The excitation efficiency of acoustic phonons was significantly lower with single pulses and coherent excitation was made possible by using a continuous train of pulses with a repetition frequency close to the target phonon resonance. As shown in Fig. 2 (c), where the red line is the spectrum of the BWD optical wave and the black dotted line is the Rayleigh scattering alone that serves as a reference, the Stokes component (area in red) was enhanced and the energy of the higher frequency components was transferred to it. This experiment serves to verify acoustic phonons can be excited by OFCs with repetition rates equal to the phonon resonance.

To demonstrate coherent control of acoustic phonons, the repetition rate of the OFC was changed to 7.8 GHz (half the phonon resonance frequency), and a Michelson interferometer

(MI) was inserted to produce double pulses as shown in Fig. 2 (d) (see Method for further details). Here, the pulse delay was normalized to the inverse of half the phonon resonance (128 ps). As depicted in Fig. 3 (a), the phonon amplitude can be accumulated when the interval of a series of optical pulses matches the phonon resonance, which is identical to the results of the previous experiment illustrated in Fig. 2. In contrast, when the delay is set to the inverse of 0.25 or 0.75 of the phonon resonance frequency, while the first pulse excites the phonon, the second pulse damps it (Fig. 3 (b)). As a result, the amplitude of the phonon is suppressed even if the fundamental repetition frequency of the OFC is matched to the phonon resonance frequency.

In the experiment, by sweeping the length of one of the interferometer's arm via a motorized stage, double pulses with a normalized time delay ranging from 0 to 1.25 of the inverse of the phonon resonance frequency could be obtained. Figure 3 (c) shows the phonon amplitude as a function of this normalized delay. The red trace is for the on-resonance case ($\omega_{rep} = \omega_{AP}/2$) and the blue trace represents the out–of-resonance case ($\omega_{rep} = \frac{\omega_{AP}}{2} + 2\pi \times 34$ MHz). The red trace exhibits a strong dependence with the time interval of the double pulse. When the interval was set to 0, 0.5 or 1, the phonon amplitude increased. On the other hand, the phonon amplitude was suppressed when the delay was set to 0.25 and 0.75. The dashed line in Fig 3 (c) shows the simulated result, which is in good agreement with the experimental results. Figure 3 (d) and (e) show the optical spectra of the BWD light as the function of the normalized pulse interval ((d) at resonance and (e) out of resonance). The Stokes components in Fig. 3 (d) are enhanced when the pulse interval is 0, 0.5 or 1, which supports the concept as shown in Fig. 1 (e) and (f). In contrast, when the phonon is out of resonance, the optical spectrum does not change even though the pulse interval changes (Fig. 3 (e)).

In conclusion, we demonstrate coherent control of acoustic phonons in a standard single-mode fiber by a repetition-frequency-tailored multi-GHz OFC. Not only was the phonon amplitude coherently enhanced, but the acoustic phonon was also controlled by the optical pulse delay. Our new approach to control the effective interaction strength between photons and phonons, can be applied in various areas of science, especially in the field of telecommunications. In this letter, we focus on acoustic phonon experiments, but the result opens an entirely new optical scheme to manipulate (quasi) particles in the field of solid-state physics, where there are a myriad of intriguing elemental excitations such as optically accessible skyrmions[20] and magnons[21,22]. The technique could be extended to controlling multiple kinds of excitations simultaneously by combining other degrees of freedom of the OFC, such as the carrier-envelope offset frequency, the pulse duration or the chirp, to name a few. We believe applying OFCs to study and manipulate particles has great potential in the field of solid-state physics and that our work represents an important step in "comb-matter interactions" research.

**Methods**

*The light source (multi-GHz OFC)*

An OFC based on a Kerr-lens mode-locked Yb:$Y_2O_3$ ceramic laser with a repetition rate of 7.8/15.6 GHz was used. The center wavelength and pulse duration were 1080 nm (278 THz) and 152 fs, respectively. The repetition rate was stabilized to an analog signal generator via a phase-locked loop circuit and a piezo actuator mounted on one of the cavity mirrors. Other details of this OFC can be found in the reference[18]. Free-running fluctuations of the carrier-envelop offset and optical frequencies were negligible, and we did not apply any stabilization to them during the experiments.

The OFC output was spectrally filtered by a hand-made optical band-pass filter (OBPF) with a maximum frequency resolution of 2 GHz at the wavelength of 1080 nm. It is based on a multi-pass, high-resolution spectrograph[23] and it would let through four longitudinal modes, as shown in Fig. 2. (c). The filtered output was amplified by a three-stage optical amplifier (a cascade of one semiconductor optical amplifier followed by two Yb:fiber amplifiers) to a maximum average power of more than 100 mW. The output power could be tuned by a variable optical attenuator assembled in-house.

*Preparation of the double pulse*

The Michelson interferometer (MI) was used to generate double pulses. In one of the interferometer arms, a stepper-motorized mechanical stage was used to tune the pulse time interval between 0 and 160 ps. A piezo actuator and a noise source were used to scan the arm length in order to average out the spectral interference of the double pulse to achieve higher SNR. Without this piezo actuator, sub-wavelength stabilization of the interferometer's arm would be required, and the cost would be lengthy scan times and data averaging. While the use of a piezo actuator reduces the time resolution of the experiment to several hundreds of femtoseconds, such a high time resolution was not required for our measurements. The peak-to-peak amplitude and the bandwidth were measured to be 3 µm and 300 Hz, respectively.

*Photon-phonon interaction*

The field distribution of the FWD and BWD optical modes in the fiber are expressed as

$$E_{FWD}(z,t) = \sum_n A_{FWD}^{(n)}(z,t) \exp\left[i\left(k_{FWD}^{(n)}z - \omega_{FWD}^{(n)}t\right)\right] + \text{c.c.}$$

$$E_{BWD}(z,t) = \sum_n A_{BWD}^{(n)}(z,t) \exp\left[i\left(-k_{BWD}^{(n)}z - \omega_{BWD}^{(n)}t\right)\right] + \text{c.c.}$$

where, $A_{FWD}^{(n)}(z,t)$ and $A_{BWD}^{(n)}(z,t)$ are the amplitudes; $k_{FWD}^{(n)} = \frac{n_{FUT}\omega_{FWD}^{(n)}}{c}$, $k_{BWD}^{(n)} = \frac{n_{FUT}\omega_{BWD}^{(n)}}{c}$ are the wave numbers; c.c. represents the complex conjugate.

The intensity of each field is defined as

$$I_{FWD}^{(n)} := 2n_{FUT}\epsilon_0 c A_{FWD}^{(n)}(z,t)\left(A_{FWD}^{(n)}(z,t)\right)^*$$

$$I_{BWD}^{(n)} := 2n_{FUT}\epsilon_0 c A_{BWD}^{(n)}(z,t)\left(A_{BWD}^{(n)}(z,t)\right)^*.$$

Then the following coupling differential equations are obtained to describe the photon-phonon interaction.

$$\frac{dI_{FWD}^{(n)}}{dz} = gI_{FWD}^{(n)}I_{BWD}^{(n+1)} - gI_{FWD}^{(n)}I_{BWD}^{(n-1)} - \alpha I_{FWD}^{(n)}$$

$$-\frac{dI_{BWD}^{(n)}}{dz} = gI_{BWD}^{(n)}I_{FWD}^{(n+1)} - gI_{BWD}^{(n)}I_{FWD}^{(n-1)} - \alpha I_{BWD}^{(n)}$$

$$\rho(z,t) = \epsilon_0\gamma_e q^2 \sum_n \frac{A_{FWD}^{(n)}\left(A_{BWD}^{(n+1)}\right)^*}{\Omega_B^2 - \omega_{rep}^2 - i\omega_{rep}\Gamma_B},$$

where $\rho(z,t)$ is the intensity of the FWD acoustic wave; $\gamma_e$ is the electrostrictive constant; $q \sim 2k_{FWD}^{(1)}$ is the wave number of the FWD acoustic wave; $\Gamma_B$ is the Brillouin linewidth; $\Omega_B$ is the Brillouin frequency shift. In the steady state, the intensity of the acoustic phonon $\rho(z,t)$ can be estimated from $A_{FWD}^{(n)}$ and $A_{BWD}^{(n)}$. To calculate the simulated dashed line in Fig. 3(c), we solved these differential equations in MATLAB and as described in the supplemental information of reference[12].

**Acknowledgements**

The authors would like to thank Isao Ito for loaning experimental equipment and Dr. Alissa Silva for careful proofreading. This research was supported by the Photon Frontier Network Program of the Ministry of Education, Culture, Sports, Science and Technology (MEXT). M.E. and S.K. were supported by Grant-in-Aid for the Japan Society for the Promotion of Science (JSPS) Fellows (DC1).

**Author contributions**

M.E. is responsible for the overall execution of the experiment. M.E. and S.K. developed the laser source. M.E. and S.T. developed analytical models to explain the measured data with the assistance of Y.K. All authors participated in discussions of the results and preparation of the manuscript.

**Competing financial interests**

The authors declare no competing financial interests.

**Author information**

Correspondence and requests for the materials should be addressed to Y.K. (yohei@issp.u-tokyo.ac.jp).

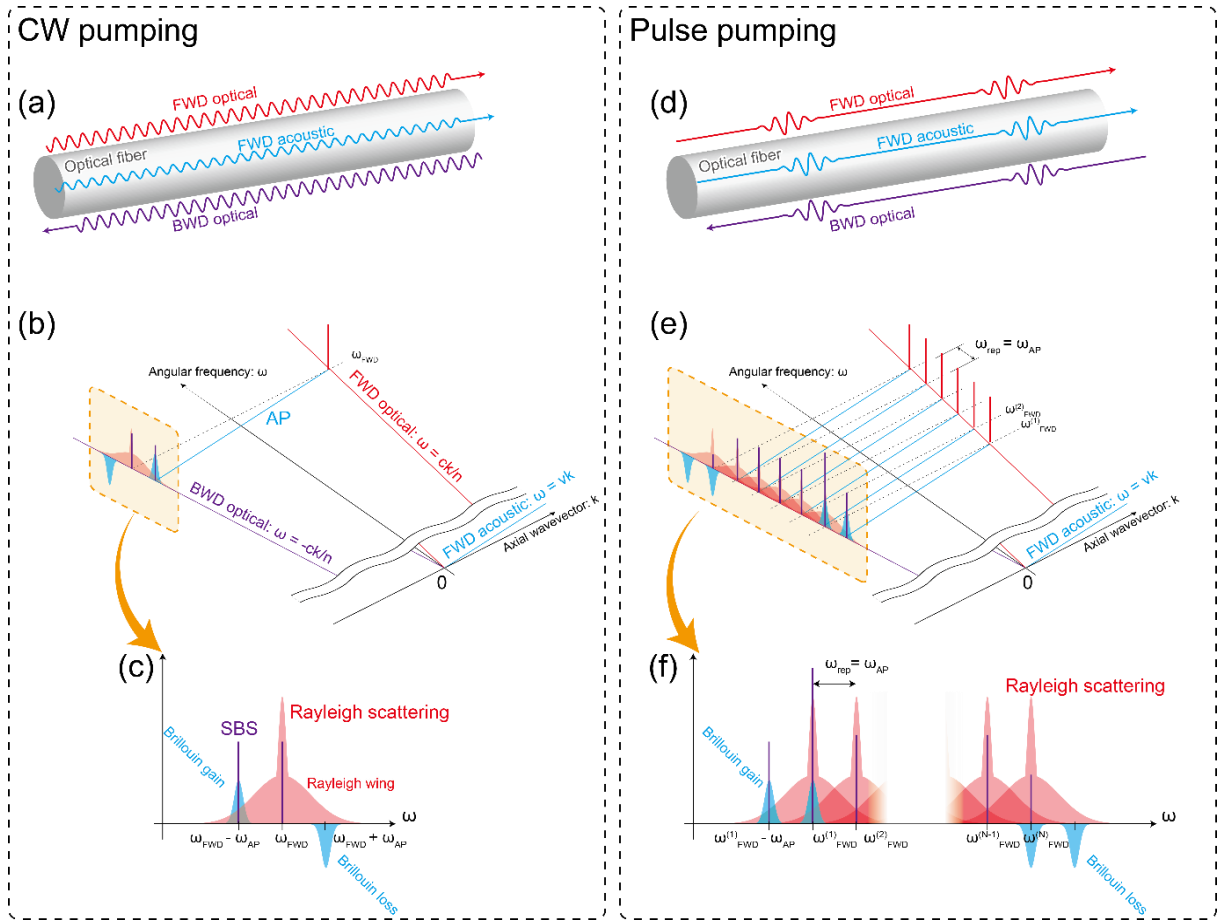

**Figure 1| Conceptual images for Brillouin scattering in a fiber and dispersion diagrams (not to scale) of the light and the acoustic waves.** In the case of CW pumping: (a), (b) and (c) are the in-fiber illustration, the dispersion diagram and the BWD optical spectrum, respectively. In the case of pulsed (multi-mode) pumping: (d), (e) and (f) are the in-fiber illustration, the dispersion diagram and the BWD optical spectrum, respectively. Red line: FWD optical waves or longitudinal modes, purple line: BWD optical waves or longitudinal modes, Sky blue line: FWD acoustic waves or longitudinal modes. In (b), (c), (e) and (f), areas indicated in red are the Rayleigh scattering components with Rayleigh wings and the areas in sky-blue designate Brillouin gain(s) and loss(es) due to the FWD acoustic waves.

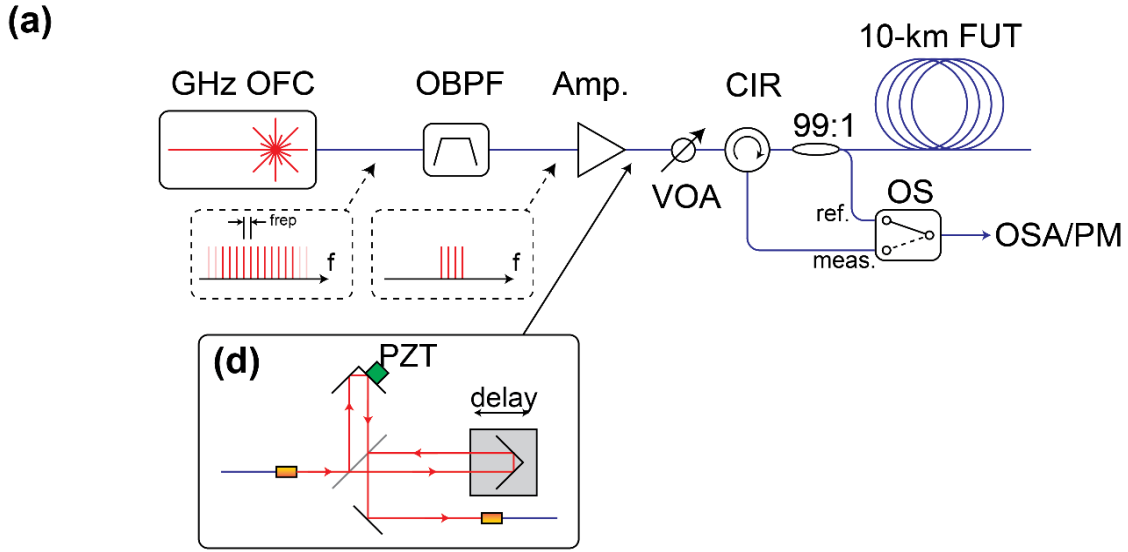

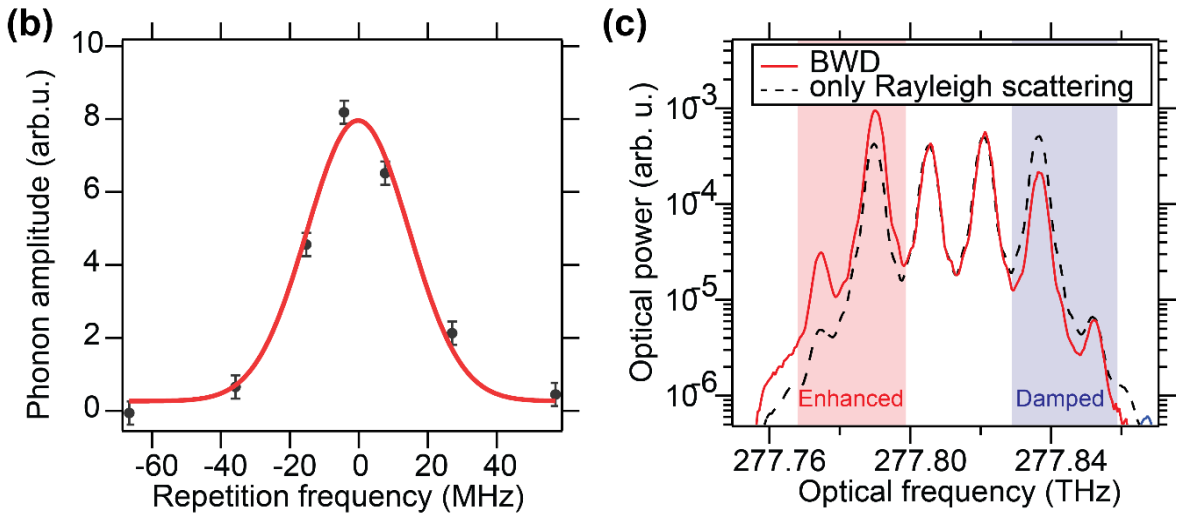

**Figure 2| Experimental apparatus and results.** (a) The experimental apparatus. KLM laser: Yb:Y$_2$O$_3$ ceramic Kerr-lens mode-locked laser with the repetition rate of 7.8 or 15.6 GHz. OBPF: optical band-pass filter based on a grating spectrometer with a resolution of 2 GHz. Amp.: Three-stage optical amplifier (made of one semiconductor optical amplifier and two ytterbium-doped fiber amplifiers). VOA: variable optical attenuator. CIR: optical circulator. OS: optical switch. FUT: fiber-under-test. OSA: optical spectrum analyzer with a resolution of 4 GHz. PM: optical power meter. (b) BWD power (corresponding to the phonon amplitude) as a function of the repetition rate detuning measured with an input power of 20 mW. The red trace is the Gaussian fit to the black experimental data points with error bars. (c) BWD spectra at the repetition rate of 15.57 GHz. Black-dashed line: Rayleigh scattering used as a reference, Red solid line: BWD spectrum. The red and blue shaded areas correspond to the Brillouin gain and loss of Fig. 1 (f), respectively. (d) the Michelson interferometer used in the double pulse experiment. PZT: piezo actuator used for scanning the arm length.

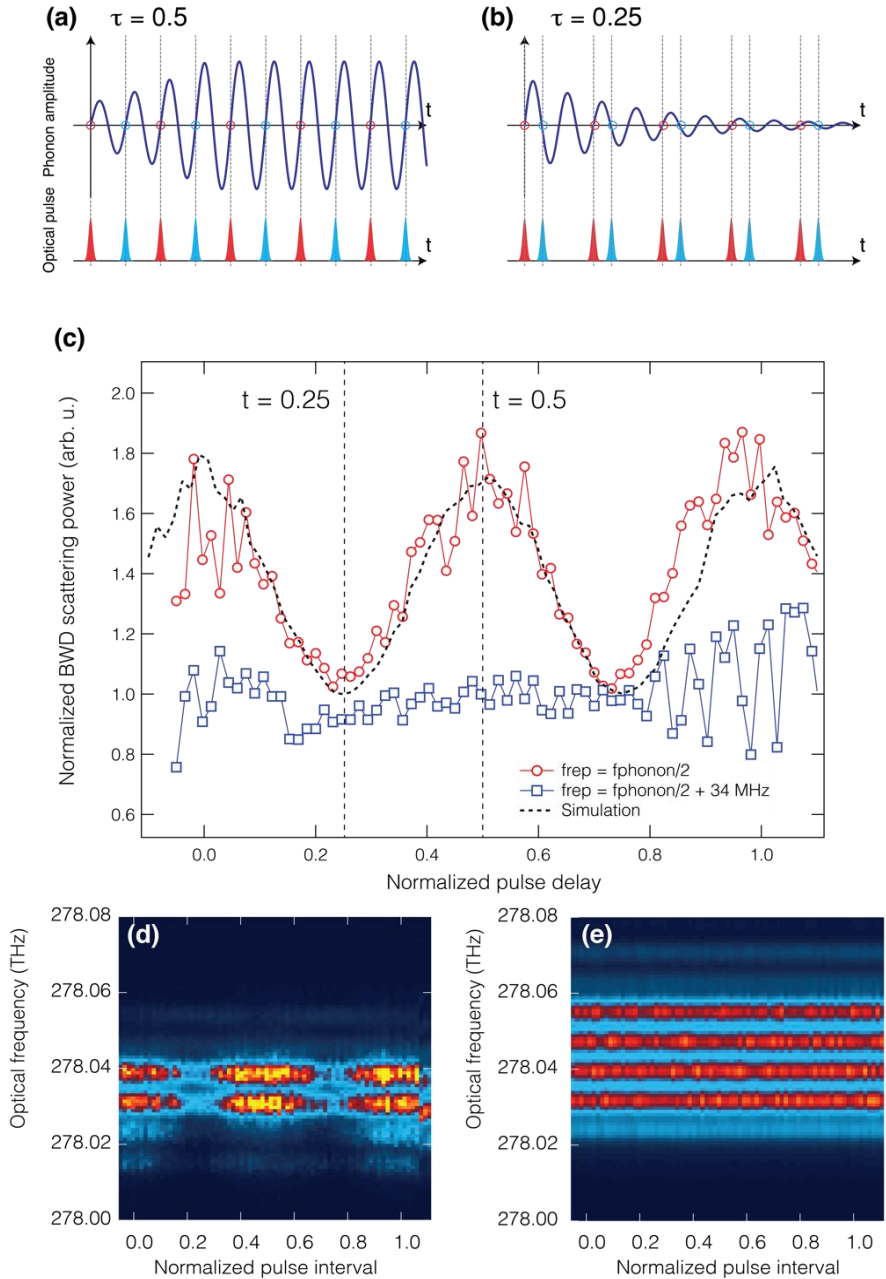

**Figure 3| Conceptual images and results of the coherent control of phonons with double pulses.** Time domain pictures for (a) $\tau = 0.5$ and (b) 0.25 delays, respectively. (a) Growth of the phonon amplitude along the optical pulse train. In contrast, (b) the phonon amplitude which is excited by the red pulses and damped by the blue ones. (c) BWD powers (corresponding to the phonon amplitude) as a function of the normalized pulse delay $\tau$. Solid red and blue traces show the experimental data with the repetition angular frequency of $\omega_{AP}/2$ and $\frac{\omega_{AP}}{2} + 2\pi \times 34$ MHz, respectively. The dashed line shows the simulation result. (d) and (e) show the optical spectra of the BWD lights when at resonance and out of resonance, respectively.